\documentclass{emulateapj}
\usepackage{graphicx}
\usepackage{mathrsfs}	
\usepackage{amsmath}
\usepackage{color}

\shorttitle{SLACS. XIII.}
\shortauthors{Shu et al. 2017}
 
\begin{document}
 

\title{The Sloan Lens ACS Survey. XIII. Discovery of 40 New Galaxy-Scale Strong Lenses\text{*}}
\altaffiltext{\text{*}}{Based on observations made with the NASA/ESA Hubble Space Telescope (\textsl{HST}), obtained at the Space Telescope Science Institute, which is operated by AURA, Inc., under NASA contract NAS 5-26555. These observations are associated with \textsl{HST} program \#12210.}

\author{Yiping Shu}
\affil{Purple Mountain Observatory, Chinese Academy of Sciences, 2 West Beijing Road, Nanjing 210008, China ({\tt yiping.shu@pmo.ac.cn})}

\author{Joel R. Brownstein}
\affil{Department of Physics and Astronomy, University of Utah, 115 South 1400 East, Salt Lake City, UT 84112, USA}

\author{Adam S. Bolton}
\affil{National Optical Astronomy Observatory, 950 North Cherry Avenue, Tucson, AZ 85719, USA}

\author{L\'eon V. E. Koopmans}
\affil{Kapteyn Astronomical Institute, University of Groningen, P.O. Box 800, NL-9700 AV Groningen, the Netherlands}

\author{Tommaso Treu}
\affil{Department of Physics and Astronomy, University of California, Los Angeles, CA 90095, USA}

\author{Antonio D. Montero-Dorta}
\affil{Departamento de F\'isica Matem\'atica, Instituto de F\'isica, Universidade de S\~ao Paulo, Rua do Mat\~ao 1371, CEP 05508-090, S\~ao Paulo, Brazil}

\author{Matthew W. Auger}
\affil{Institute of Astronomy, University of Cambridge, Madingley Road, Cambridge CB3 0HA, UK}

\author{Oliver Czoske}
\affil{Institut f\"{u}r Astronomie, der Universit\"{a}t Wien, T\"{u}rkenschanzstra{\ss}e 17, A-1180 Wien, Austria}

\author{Rapha\"el Gavazzi}
\affil{Institut d'Astrophysique de Paris, CNRS, UMR 7095, Universit«e Pierre et Marie Curie, 98bis Bd Arago, F-75014 Paris, France}

\author{Philip J. Marshall}
\affil{Kavli Institute for Particle Astrophysics and Cosmology, Stanford University, 452 Lomita Mall, Stanford, CA 94305, USA}

\author{Leonidas A. Moustakas}
\affil{Jet Propulsion Laboratory, California Institute of Technology, MS 169-506, 4800 Oak Grove Drive, Pasadena, CA 91109, USA}

\begin{abstract}

We present the full sample of 118 galaxy-scale strong-lens candidates in the Sloan Lens ACS 
(SLACS) Survey for the Masses (S4TM) Survey, which are spectroscopically selected from 
the final data release of the Sloan Digital Sky Survey. Follow-up \textsl{Hubble Space 
Telescope} (\textsl{HST}) imaging observations confirm that 40 candidates are definite strong 
lenses with multiple lensed images. The foreground-lens galaxies are found to be early-type 
galaxies (ETGs) at redshifts 0.06--0.44, and background sources are emission-line 
galaxies at redshifts 0.22--1.29. As an extension of the SLACS Survey, the S4TM Survey is 
the first attempt to preferentially search for strong-lens systems with relatively lower 
lens masses than those in the pre-existing strong-lens samples. 
By fitting \textsl{HST} data with a singular isothermal ellipsoid model, 
we find that the total projected mass within the Einstein radius of the S4TM strong-lens 
sample ranges from $3 \times10^{10} M_{\odot}$ to $2 \times10^{11} M_{\odot}$. 
In \citeauthor{Shu15}, we have derived the total stellar mass of the S4TM lenses to be 
$5 \times10^{10} M_{\odot}$ to $1 \times10^{12} M_{\odot}$. Both the total enclosed mass and 
stellar mass of the S4TM lenses are on average almost a factor of 2 smaller than those of 
the SLACS lenses, which also the represent typical mass scales of the current strong-lens samples. 
The extended mass coverage provided by the S4TM sample can enable a direct test, with the aid of 
strong lensing, for transitions in scaling relations, kinematic properties, mass structure, 
and dark-matter content trends of ETGs at intermediate-mass scales as noted in previous studies. 

\end{abstract}



\section{Introduction}

Early-type galaxies (ETGs) are a group of galaxies that have regular ellipsoidal shapes, 
typically old stellar populations, and little ongoing star-formation activity. They are 
believed to be the end product of a hierarchical merging scenario of galaxy formation 
\citep[e.g.,][]{TT72, White91, Kauffmann93, Cole2000}. Early works suggested that ETGs 
seemed to be a well-defined population by tightly following several empirical scaling relations 
\citep[e.g.,][]{Faber76, Kormendy77, Dressler87, Djorgovski87}. However, as the sample became 
larger and more complete later on, clear transitions in several scaling relations, 
kinematic properties, and dark-matter content trends of ETGs were noted at two 
characteristic mass scales, $3 \times 10^{10} M_{\odot}$ and $2 \times 10^{11} M_{\odot}$ 
\citep[e.g.,][]{Tremblay96, Graham03a, Kauffmann03, Graham08, Hyde09, Skelton09, Tortora09, 
vanderWel09, Bernardi11a, Bernardi11b, Cappellari13a, Cappellari13, Montero16}. 
This implies that that physical processes that regulate how ETGs form and evolve must have 
undergone similar transitions at these two mass scales. 

Previous studies on the ETG transitions primarily used photometric data or stellar kinematics 
data inferred from spectra for ETG mass estimation, which are known to be model dependent and 
have weak constraining power on the dark-matter content. 
The strong gravitational lensing phenomenon, which is the appearance of multiple images of the 
same distant source due to the convergence of light caused by the gravitational field of an 
intervening object (denoted as the ``lens'') as a prediction of Albert Einstein's general 
relativity \citep[GR;][]{Einstein16}, provides a robust way of determining 
the total mass in the central region of the lens object \citep[e.g., see a review article by][]{Treu10}. 
Accurate mass measurements of ETG lens systems may provide new 
insights in understanding of such transitions, especially by combining low-, intermediate-, 
and  high-mass ETG strong-lens samples. 

Nevertheless, strong-lensing events are rare because it requires a close alignment among 
the observer, the lens, and the source. 
The probability of a lensing event occurring is characterized by the lensing cross section, 
which is the area on the source plane within which the source needs to be to 
produce at least two images. To the leading order, the lensing cross section is determined by 
the mass of the lens object, at least on the galaxy-scales that we are considering in this paper. 
Because of this, current galaxy-scale strong-lens searches are strongly biased toward massive 
ETGs for high success-rates. Over the past four decades, the number of strong-lens systems has 
accumulated to just a few hundred\footnote{Number based on the Master Lens Database 
(http://admin.masterlens.org/index.php?)} from dedicated photometric and/or spectroscopic 
surveys 
\citep[e.g.,][]{Walsh79, Munoz98, Kochanek00, Browne03, Ebeling07, SLACSV, Faure08, SWELLSI, 
Brownstein12, More12, Inada12, Sonnenfeld13, Stark13, Vieira13, Pawase14, More16a, BELLS_IV, 
Negrello17, Sonnenfeld17}. The typical stellar mass of the current galaxy-scale strong-lens 
sample is several times $10^{11} M_{\odot}$ \citep[e.g.,][]{SLACSX, Faure11, Brownstein12, 
Sonnenfeld13}, beyond the above-mentioned characteristic mass scales. 
Note that this mass peak is primarily the result of the lensing cross section per lens, which 
is proportional to the mass to the second power, and galaxy mass function, which suggests that the 
number of ETGs typically increases with $1/M$ below the characteristic mass $M_*$ and declines 
exponentially beyond \citep[e.g.,][]{Li09, Yang09, Ilbert10, Baldry12, Maraston13, Davidzon17}. 
Clearly, a large sample of strong-lens systems containing low- and intermediate-mass 
ETG lenses is needed. 

The Sloan Lens ACS (SLACS) Survey for the Masses (S4TM) Survey has been designed as an 
attempt to preferentially select relatively lower-mass strong-lens systems. 
To achieve that, we rely on the most prolific strong-lens selection technique ever developed, 
the one presented in \citet{Bolton04}. This technique has lead to several major strong-lens 
surveys including the Sloan Lens ACS Survey \citep[SLACS;][]{SLACSV, SLACSIX}, 
the Sloan WFC Edge-on Late-type Lens Survey \citep[SWELLS;][]{SWELLSI, Brewer12}, 
the BOSS Emission-Line Lens Survey \citep[BELLS;][]{Brownstein12}, 
and the BELLS for GAlaxy-Ly$\alpha$ EmitteR sYstems survey \citep[BELLS GALLERY;][]{BELLS_IV}. 
From \textsl{Hubble Space Telescope} (\textsl{HST}) follow-up imaging observations, 
we have already confirmed nearly 150 strong-lens systems in total (85 in SLACS, 20 in SWELLS, 
25 in BELLS, and 17 in BELLS GALLERY). However, previously, candidates with the highest 
 predicted lensing cross sections (essentially largest lens masses) were prioritized in these 
\textsl{HST} observations. In the S4TM Survey, we try to extend the lens-mass 
coverage by targeting at candidates with relatively lower predicted lens mass 
at the cost of lowering the success rate. 
We will explain how we achieve this in Section~\ref{sect:sample}. 

This paper is organized as follows. Section~\ref{sect:sample} briefly describes how the 
lower-mass S4TM sample is selected. \textsl{HST} photometric data and strong-lensing analysis 
are provided in Sections~\ref{sect:photometry} and \ref{sect:models}. 
Section~\ref{sect:discussion} presents the discussion followed by a summary in 
Section~\ref{sect:summary}. Throughout the paper, 
we adopt a cosmological model with $\rm \Omega_m = 0.274$, 
$\rm \Omega_{\Lambda} = 0.726$ and $H_0 \rm = 70\,km\,s^{-1}\,Mpc^{-1}$ 
\citep[WMAP7;][]{WMAP7}.

\section{Sample Selection}
\label{sect:sample}

As an extension of the SLACS survey, the S4TM survey selects 
strong-lens candidates spectroscopically from the galaxy-spectrum database provided 
by the seventh and final data release of the Sloan Digital Sky Survey \citep[SDSS;][]{DR7}. 
The S4TM survey adopts the strong-lens selection technique 
that led to the successful discoveries of nearly 150 strong-lens systems 
\citep[e.g.,][]{SLACSV, SLACSIX, SWELLSI, Brewer12, Brownstein12, BELLS_IV, Rui17}. 
The underlying principle is to select the candidate that shows multiple nebular emission 
lines in their spectra, collected by an optical fiber at a common redshift that is 
significantly higher than the candidate itself. Such a special configuration indicates 
that there are two objects at different redshifts within the same light cone, which is 
usually as narrow as 2--3 arcsec in diameter, and a lensing event is likely to 
happen. High-resolution follow-up imaging observations are then obtained to confirm 
the lensing nature of the system. 
More detailed descriptions on this technique can be found in 
\citet{Bolton04}, \citet{Brownstein12}, and \citet{BELLS_IV}. 

After picking out strong-lens candidates with higher-redshift nebular emission lines 
from the SDSS DR7 database, we first perform a morphology cut by only retaining candidates 
with early-type morphology as determined from the SDSS images. Then we compute an approximate 
strong-lensing Einstein radius, $\theta_{\rm Ein}$, based on the foreground and background 
redshifts and SDSS measured central stellar velocity dispersion, assuming a singular 
isothermal sphere model. As shown in \citet{SLACSV}, the SLACS lens confirmation 
rate is an increasing function of $\theta_{\rm Ein}$. We remove candidates with 
a predicted $\theta_{\rm Ein}$ smaller than 0\farcs5 because the confirmation rate drops 
rapidly to $\leqslant 10\%$ below this angular scale \citep{SLACSV}. 

To specifically select lens galaxies to complement the SLACS survey in terms of lens-galaxy 
mass distribution, we rely on a dimensional mass variable defined as 
$M_{\rm dim} = G^{-1} \sigma^2 R_{\rm eff}/2$ where $\sigma$ and $R_{\rm eff}$ are the SDSS 
measured stellar velocity dispersion and effective radius, respectively. Directly constructed 
from existing SDSS measurements, this dimensional mass, $M_{\rm dim}$, serves as a simple 
gauge of the lens galaxy mass, at least in a relative sense. 
Candidates with $M_{\rm dim}$ less than $10^{10.5} M_{\odot}$ 
($\approx 3 \times 10^{10} M_{\odot}$), a mass scale below which is sparsely populated by 
confirmed SLACS lenses, are included into the S4TM sample. 
An analysis of the SLACS sample further shows that the ratio of Einstein radius to effective 
radius of SLACS lenses is limited to a range from $\sim$0.4 to 1.0. 
This ratio is a useful scale in galaxy-scale strong lenses. Although mass measurements 
inferred from strong lensing are extremely accurate, they are limited to a physical radial 
aperture --- the Einstein radius --- that is determined by serendipitous cosmic geometry. 
In order to control effectively for systematic mass-aperture effects in the follow-up 
lensing and dynamical analyses, we would like to build up ensembles of strong-lens systems 
with a significant variation in the ratio of Einstein radius to optical effective radius 
for multiple fixed lens-galaxy mass ranges. Here the effective radius is a normalization 
factor. As a result, we also include in the S4TM sample candidates with similar dimensional 
masses as the SLACS lenses ($M_{\rm dim} > 10^{10.5} M_{\odot}$), but with a predicted ratio 
of Einstein radius to effective radius $\theta_{\rm Ein}/R_{\rm eff}$ either less than 0.4 
or greater than 1.0. 

Eventually, the S4TM sample comprises 135 new galaxy--galaxy lens candidates. In combination 
with the SLACS sample, this lens ensemble covers nearly two decades in mass, with dense 
mapping of enclosed mass as a function of radius out to the effective radius and beyond. 

\section{\textsl{HST} Photometric Data}
\label{sect:photometry}

\begin{figure*}[htbp]
	\includegraphics[width=0.95\textwidth]{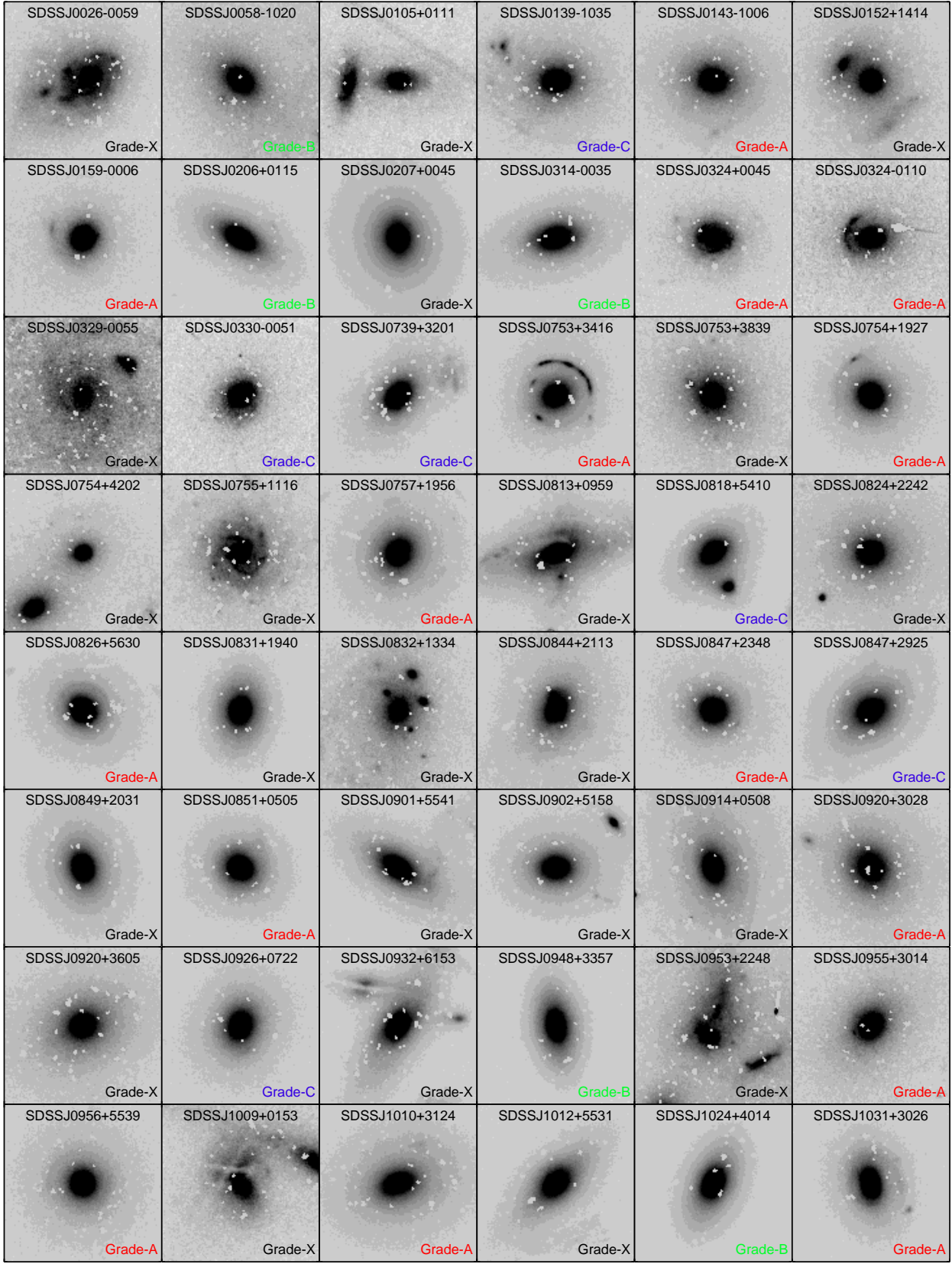}
	\caption{\label{fig:mosaic} Mosaic of the \textsl{HST} F814W-band images of the 118 S4TM strong-lens candidates. Images are $6^{\prime \prime} \times 6^{\prime \prime}$ with north up and east to the left. The SDSS name and lens grade are given for each system.}
\end{figure*}
\addtocounter{figure}{-1}
\begin{figure*}[htbp]
	\includegraphics[width=0.95\textwidth]{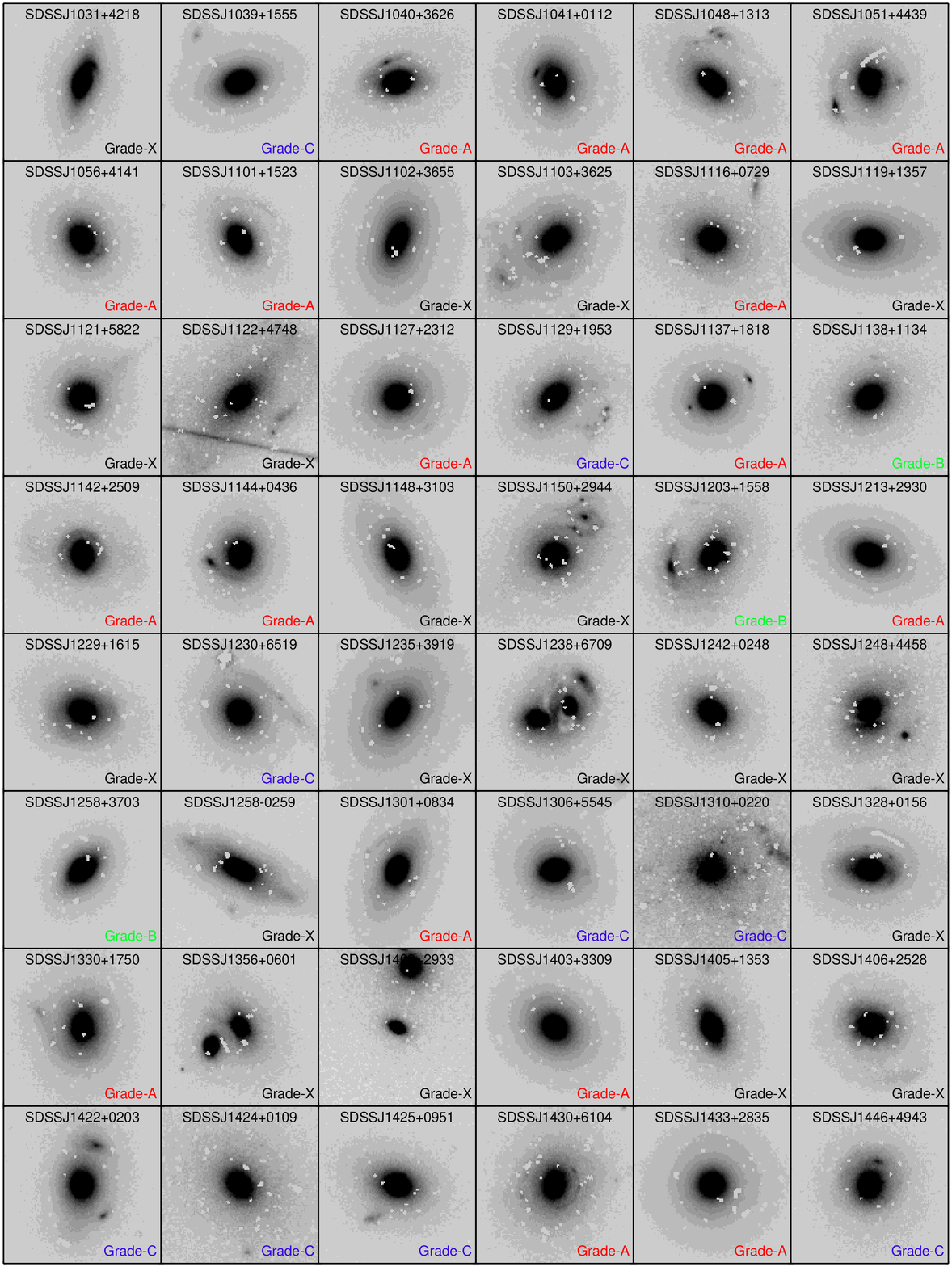}
	\caption{\textit{Continued}}
\end{figure*}
\addtocounter{figure}{-1}
\begin{figure*}[htbp]
	\includegraphics[width=0.95\textwidth]{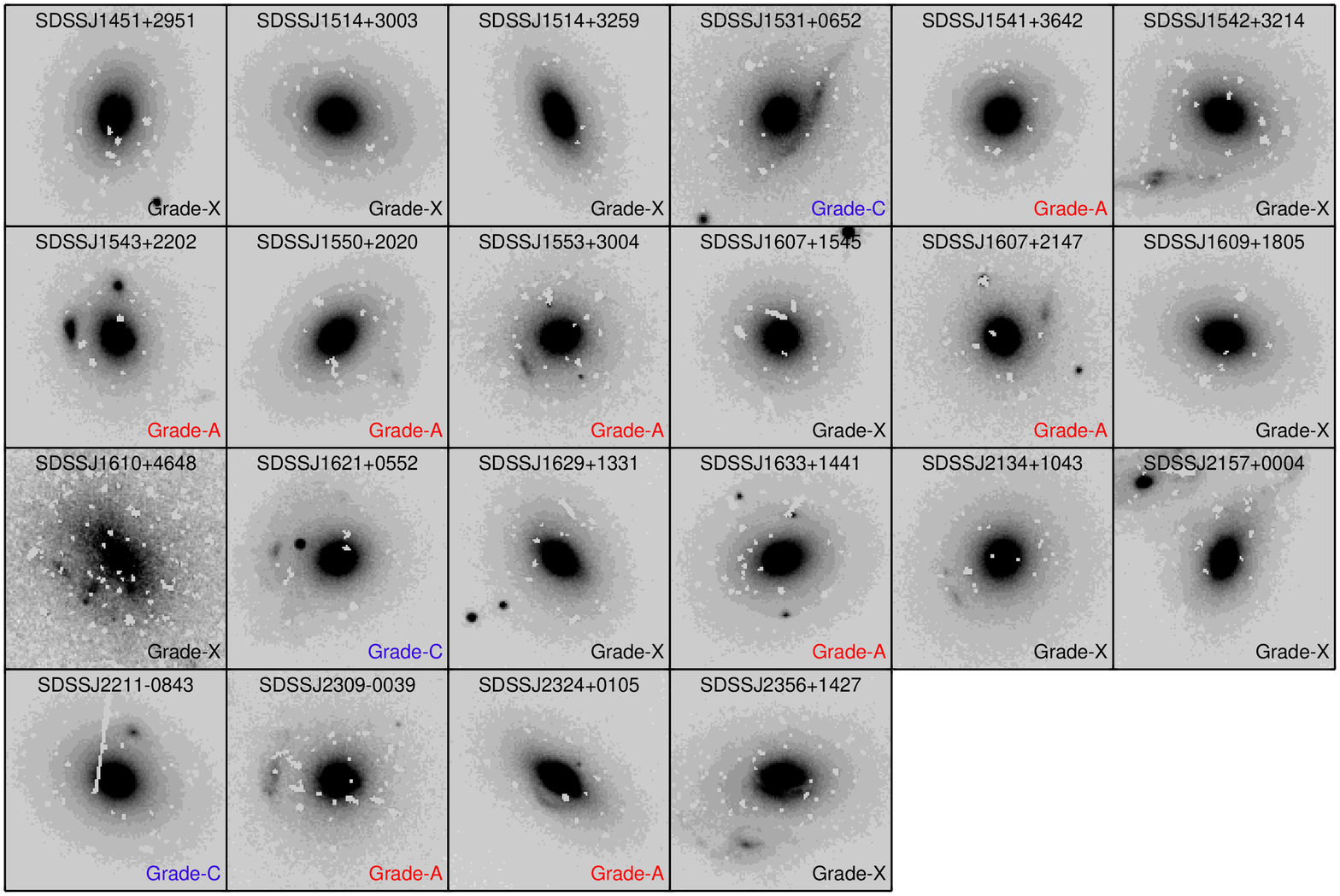}
	\caption{\textit{Continued}}
\end{figure*}

\textsl{HST} imaging observations of the S4TM sample were carried out in the F814W-band with 
the Wide Field Channel (WFC) of the ACS camera under the Snapshot Program 
\#12210 in Cycle 18 (PI: A. Bolton). Each candidate is designed to have a single exposure 
of 420 s during one \textsl{HST} visit. As of its completion, 118 visits are successfully 
observed, 2 visits are not usable (29, 35) because of guide star acquisition failure, 
and 15 visits are withdrawn. From now on, we will only focus on the 118 candidates with 
\textsl{HST} observations. 
The individual fully-calibrated, flat-fielded (FLT) files are downloaded from the \textsl{HST} 
archive and reduced by our custom-built tool, {\tt ACSPROC}, presented in 
\citet{Brownstein12}. 
In order to be consistent with \citet{SLACSV}, \citet{Brownstein12}, 
and especially \citet{Shu15}, which presents the first scientific results of the S4TM survey, 
we model the foreground lens-galaxy light with an elliptical radial B-spline model 
\citep{SLACSI}. Besides a B-spline model, we also fit the two-dimensional elliptical 
de Vaucouleurs model \citep{deVaucouleurs48} 
to the foreground light to derive some standardized quantities such as the effective radius, 
axis ratio, and major-axis position angle. Such values along with other useful information 
determined from the SDSS spectroscopic data are presented in Table~\ref{tb:tb1}. 

The B-spline-subtracted residual images are inspected by a group of authors (A.S.B., 
J.R.B., and Y.S.) to determine the lens morphology, multiplicity, and lens grade. 
As shown by the classification codes in Table~\ref{tb:tb1}, the majority of the 118 
candidates contain single ETGs in the foreground. 
As mentioned in \citet{Shu15}, we identify 40 grade-A strong lenses with 
definite multiple lensed images, 8 grade-B systems with strong evidence of multiple images 
but insufficient signal-to-noise ratio for definite conclusion and/or modeling, 
and 18 grade-C systems clearly showing lensed images of the background galaxies but no clear 
counterimages. The remaining 52 candidates are classified as nonlenses (grade-X). 
Figure~\ref{fig:mosaic} shows the mosaic of the fully-reduced \textsl{HST} F814W-band images 
of all of the 118 systems. Target names and lens grades are given in each 
$6^{\prime \prime} \times 6^{\prime \prime}$ stamp. The small white blocks in each stamp 
correspond to the pixels masked due to cosmic rays, which we can not correct for 
based on single exposures. 
The success rate of finding grade-A lenses in the S4TM survey ($34\%$) is 
slightly lower than those in previous SLACS and BELLS surveys, which can reach about 50\%. 
That could be related to the trade-off between success rate and lens-mass coverage in the 
S4TM Survey as discussed above. 
The average lens and source redshifts are 0.17 and 0.61, slightly lower than those of 
the SLACS lens sample (0.21 and 0.63, respectively). We note that none of the four galaxies 
in Table~\ref{tb:tb1} with zero stellar velocity dispersions according to the SDSS 
reduction pipeline are grade-A.  

\section{Strong-lensing Analysis}
\label{sect:models}

Here we only report strong-lens modeling results for S4TM grade-A lenses, and refer the 
interested reader to \citet{Shu15} for the results of the S4TM grade-C lenses and a combined 
analysis of grade-A and grade-C lenses in the SLACS and S4TM surveys. 

Lens modeling is done with the custom-built tool {\tt lfit\_gui} first introduced in 
\citet{BELLS_IV}. There are three components in the lens model. 
The first component is the foreground-light model. 
For the same consistency reason, we adopt the B-spline fit as the model for the foreground-light 
distribution following \citet{Shu15}. Note that this foreground-subtraction strategy could 
introduce some systematic uncertainties in the lens and source parameters as discussed in 
\citet{Marshall07} and \citet{Shu16a, BELLS_IV}. Following our previous works, 
we use the singular isothermal ellipsoid (SIE) profile to model the projected lens-mass 
distribution. Our singular isothermal ellipsoid (SIE) model has a two-dimensional surface 
mass--density profile following \citet{Kormann94} as 
\begin{equation}
\Sigma (x, y) = \Sigma_{\rm crit} \frac{\sqrt{q}}{2} \frac{b_{\rm SIE}}{\sqrt{x^2+q^2 y^2}},
\end{equation}
where $\Sigma_{\rm crit}$ is the critical density determined by the cosmological 
distances as 
\begin{equation}
\Sigma_{\rm crit} = \frac{c^2}{4 \pi G} \frac{d_{S}}{d_{L} d_{LS}}, 
\end{equation}
and $d_{L}$, $d_{S}$, and $d_{LS}$ are the angular diameter distances 
from the observer to the lens, from the observer to the source, 
and between the lens and the source, respectively. 
We do not include any external shear in the lens model because it is a minor 
effect as quantified in \citet{Shu15}. 
The last component is the source model. As explained in \citet{BELLS_IV}, {\tt lfit\_gui} 
provides two types of source models. One is the parametric source model in which the 
source-light distribution is characterized by multiple elliptical S\'{e}rsic components. 
The other is the pixelized source model obtained from a direct inversion 
\citep[e.g.,][]{Dye05, Koopmans05, Brewer06, Suyu06, Vegetti09, Nightingale15}. We start from 
a single S\'{e}rsic source component, and then generate a pixelized source model with 
all the lens model parameters fixed. Extra S\'{e}rsic components are added to match the 
pixelized source model. This procedure is done iteratively until the parametric source model 
and the pixelized source model are in reasonable agreement. 
The parameter optimization is done by minimizing a $\chi^2$ function using the 
Levenberg--Marquardt algorithm as implemented in the LMFIT package \citep{lmfit}. 

\begin{table*}[htbp]
\begin{center}
\caption{\label{tb:tb1} Selected properties of the S4TM sample.}
\begin{tabular}{c c c c c c c c c c c}
\hline \hline
Target & Plate-MJD-Fiber & $z_{L}$ & $z_{S}$ & $\sigma_{\rm SDSS}$ & $I_{\rm 814}$ & $\Delta I_{\rm 814}$ & $R_{\rm eff}$ & $q$ & P.A. & Classification \\
 & & & & (km\,s$^{-1}$) & (mag) & (mag) & (arcsec) & & (deg) & \\
(1) & (2) & (3) & (4) & (5) & (6) & (7) & (8) & (9) & (10) & (11) \\
\hline
SDSSJ0026$-$0059 & 0391-51782-255 & 0.0924 & 0.9506 & 74 $\pm$   19 & 16.87 & 0.04 & 4.92 & 0.65 & 125 & E-S-X \\ 
SDSSJ0058$-$1020 & 0658-52146-191 & 0.3088 & 0.7741 & 295 $\pm$   23 & 16.76 & 0.07 & 4.76 & 0.69 & 47 & E-M-B \\ 
SDSSJ0105$+$0111 & 0670-52520-403 & 0.3584 & 1.1041 & 0 $\pm$    0 & 18.99 & 0.04 & 0.83 & 0.58 & 89 & E-S-X \\ 
SDSSJ0139$-$1035 & 0663-52145-201 & 0.2221 & 0.9745 & 209 $\pm$   13 & 17.20 & 0.04 & 2.13 & 0.83 & 96 & E-U-C \\ 
SDSSJ0143$-$1006 & 0664-52174-259 & 0.2210 & 1.1046 & 203 $\pm$   17 & 16.85 & 0.05 & 3.24 & 0.78 & 82 & E-S-A \\ 
SDSSJ0152$+$1414 & 0430-51877-473 & 0.1359 & 0.2920 & 121 $\pm$   21 & 16.93 & 0.12 & 3.52 & 0.82 & 22 & E-U-X \\ 
SDSSJ0159$-$0006 & 1555-53287-171 & 0.1584 & 0.7477 & 216 $\pm$   18 & 17.47 & 0.05 & 1.58 & 0.91 & 139 & E-S-A \\ 
SDSSJ0206$+$0115 & 0404-51877-530 & 0.1373 & 0.8749 & 187 $\pm$   12 & 16.93 & 0.05 & 1.13 & 0.52 & 58 & E-S-B \\ 
SDSSJ0207$+$0045 & 0404-51812-540 & 0.0419 & 1.1148 & 155 $\pm$    4 & 14.80 & 0.05 & 3.10 & 0.83 & 13 & E-S-X \\ 
SDSSJ0314$-$0035 & 0412-52258-030 & 0.1151 & 1.1501 & 172 $\pm$   10 & 16.92 & 0.14 & 1.44 & 0.61 & 105 & E-S-B \\ 
SDSSJ0324$+$0045 & 1629-52945-424 & 0.3210 & 0.9199 & 183 $\pm$   19 & 18.23 & 0.22 & 1.67 & 0.84 & 90 & E-S-A \\ 
SDSSJ0324$-$0110 & 1566-53003-246 & 0.4456 & 0.6239 & 310 $\pm$   38 & 18.17 & 0.17 & 2.23 & 0.73 & 90 & E-S-A \\ 
SDSSJ0329$-$0055 & 0713-52178-298 & 0.1062 & 0.6576 & 22 $\pm$   75 & 16.84 & 0.23 & 10.00 & 0.87 & 21 & L-S-X \\ 
SDSSJ0330$-$0051 & 0810-52672-304 & 0.3406 & 1.1334 & 194 $\pm$   34 & 19.05 & 0.22 & 0.55 & 0.77 & 126 & E-S-C \\ 
SDSSJ0739$+$3201 & 0541-51959-078 & 0.1860 & 0.6198 & 197 $\pm$    6 & 17.10 & 0.09 & 1.00 & 0.66 & 138 & E-S-C \\ 
SDSSJ0753$+$3416 & 0756-52577-482 & 0.1371 & 0.9628 & 208 $\pm$   12 & 16.55 & 0.10 & 1.89 & 0.86 & 137 & E-S-A \\ 
SDSSJ0753$+$3839 & 0544-52201-314 & 0.0408 & 1.2344 & 27 $\pm$   25 & 17.10 & 0.08 & 3.37 & 0.86 & 26 & E-S-X \\ 
SDSSJ0754$+$1927 & 1582-52939-627 & 0.1534 & 0.7401 & 193 $\pm$   16 & 17.02 & 0.10 & 1.46 & 0.94 & 45 & E-S-A \\ 
SDSSJ0754$+$4202 & 0434-51885-075 & 0.3692 & 1.0543 & 342 $\pm$   71 & 17.19 & 0.08 & 4.63 & 0.56 & 133 & E-S-X \\ 
SDSSJ0755$+$1116 & 2418-53794-354 & 0.1378 & 0.3448 & 0 $\pm$    0 & 17.16 & 0.05 & 3.74 & 0.99 & 110 & L-S-X \\ 
SDSSJ0757$+$1956 & 1922-53315-347 & 0.1206 & 0.8326 & 206 $\pm$   11 & 15.82 & 0.09 & 3.67 & 0.91 & 154 & E-S-A \\ 
SDSSJ0813$+$0959 & 2421-54153-171 & 0.1565 & 1.1851 & 195 $\pm$   13 & 16.36 & 0.05 & 2.33 & 0.52 & 110 & L-S-X \\ 
SDSSJ0818$+$5410 & 1782-53299-266 & 0.1163 & 0.3673 & 191 $\pm$   12 & 16.90 & 0.10 & 1.06 & 0.68 & 130 & E-S-C \\ 
SDSSJ0824$+$2242 & 1927-53321-521 & 0.2802 & 0.8457 & 321 $\pm$   24 & 16.17 & 0.08 & 6.67 & 0.93 & 78 & E-M-X \\ 
SDSSJ0826$+$5630 & 1783-53386-414 & 0.1318 & 1.2907 & 163 $\pm$    8 & 16.27 & 0.11 & 1.64 & 0.87 & 51 & E-S-A \\ 
SDSSJ0831$+$1940 & 2275-53709-362 & 0.0876 & 0.8805 & 155 $\pm$    6 & 16.87 & 0.07 & 1.26 & 0.72 & 173 & E-S-X \\ 
SDSSJ0832$+$1334 & 2425-54139-062 & 0.3968 & 0.7437 & 303 $\pm$   24 & 16.98 & 0.11 & 3.97 & 0.74 & 1 & E-M-X \\ 
SDSSJ0844$+$2113 & 2280-53680-388 & 0.1779 & 0.3091 & 246 $\pm$   15 & 16.22 & 0.07 & 3.96 & 0.74 & 162 & E-S-X \\ 
SDSSJ0847$+$2348 & 2085-53379-342 & 0.1551 & 0.5327 & 199 $\pm$   16 & 17.00 & 0.06 & 1.54 & 0.94 & 90 & E-S-A \\ 
SDSSJ0847$+$2925 & 1589-52972-252 & 0.1001 & 0.2390 & 228 $\pm$    9 & 15.73 & 0.08 & 2.19 & 0.71 & 130 & E-S-C \\ 
SDSSJ0849$+$2031 & 2280-53680-144 & 0.0844 & 0.4059 & 200 $\pm$    8 & 15.95 & 0.06 & 1.97 & 0.75 & 13 & E-S-X \\ 
SDSSJ0851$+$0505 & 1189-52668-132 & 0.1276 & 0.6371 & 175 $\pm$   11 & 16.77 & 0.11 & 1.35 & 0.90 & 52 & E-S-A \\ 
SDSSJ0901$+$5541 & 0450-51908-388 & 0.1163 & 0.2467 & 194 $\pm$   10 & 16.59 & 0.04 & 2.13 & 0.53 & 58 & E-S-X \\ 
SDSSJ0902$+$5158 & 0552-51992-466 & 0.1366 & 0.2036 & 256 $\pm$    8 & 16.24 & 0.04 & 2.25 & 0.77 & 90 & E-S-X \\ 
SDSSJ0914$+$0508 & 1193-52652-142 & 0.1355 & 0.4034 & 209 $\pm$   11 & 15.44 & 0.10 & 5.60 & 0.61 & 12 & L-S-X \\ 
SDSSJ0920$+$3028 & 1938-53379-111 & 0.2881 & 0.3918 & 297 $\pm$   17 & 16.25 & 0.05 & 4.25 & 0.93 & 30 & E-S-A \\ 
SDSSJ0920$+$3605 & 1274-52995-386 & 0.1844 & 0.2731 & 238 $\pm$   11 & 16.32 & 0.03 & 3.68 & 0.79 & 116 & E-S-X \\ 
SDSSJ0926$+$0722 & 1195-52724-599 & 0.0756 & 0.2855 & 170 $\pm$   10 & 16.57 & 0.10 & 1.31 & 0.81 & 161 & E-S-C \\ 
SDSSJ0932$+$6153 & 0486-51910-350 & 0.1235 & 0.2623 & 205 $\pm$   12 & 16.74 & 0.08 & 1.84 & 0.55 & 150 & E-S-X \\ 
SDSSJ0948$+$3357 & 1945-53387-560 & 0.0814 & 1.0600 & 144 $\pm$    6 & 16.63 & 0.02 & 0.65 & 0.57 & 9 & E-S-B \\ 
SDSSJ0953$+$2248 & 2295-53734-624 & 0.0761 & 0.1743 & 0 $\pm$   96 & 15.87 & 0.05 & 10.00 & 0.82 & 160 & U-S-X \\ 
SDSSJ0955$+$3014 & 1950-53436-379 & 0.3214 & 0.4671 & 271 $\pm$   33 & 17.26 & 0.04 & 2.95 & 0.72 & 140 & E-S-A \\ 
SDSSJ0956$+$5539 & 0945-52652-390 & 0.1959 & 0.8483 & 188 $\pm$   11 & 16.84 & 0.02 & 1.96 & 0.98 & 29 & E-S-A \\ 
SDSSJ1009$+$0153 & 0502-51957-235 & 0.3352 & 0.9278 & 214 $\pm$   23 & 17.23 & 0.09 & 5.09 & 0.91 & 175 & U-S-X \\ 
SDSSJ1010$+$3124 & 1952-53378-114 & 0.1668 & 0.4245 & 221 $\pm$   11 & 15.98 & 0.05 & 3.26 & 0.75 & 108 & E-S-A \\ 
SDSSJ1012$+$5531 & 0945-52652-626 & 0.1711 & 0.6973 & 203 $\pm$    8 & 16.78 & 0.01 & 1.57 & 0.55 & 136 & E-S-X \\ 
SDSSJ1024$+$4014 & 1359-53002-204 & 0.0636 & 0.3049 & 152 $\pm$    7 & 16.24 & 0.02 & 1.13 & 0.60 & 156 & E-S-B \\ 
SDSSJ1031$+$3026 & 2354-53799-403 & 0.1671 & 0.7469 & 197 $\pm$   13 & 17.01 & 0.04 & 1.04 & 0.67 & 12 & E-U-A \\ 
SDSSJ1031$+$4218 & 1360-53033-415 & 0.1193 & 0.3076 & 185 $\pm$   11 & 17.02 & 0.02 & 1.22 & 0.55 & 165 & E-S-X \\ 
SDSSJ1039$+$1555 & 2594-54177-537 & 0.0837 & 0.3236 & 194 $\pm$    5 & 15.79 & 0.06 & 1.63 & 0.68 & 106 & E-S-C \\ 
SDSSJ1040$+$3626 & 2096-53446-570 & 0.1225 & 0.2846 & 186 $\pm$   10 & 16.93 & 0.04 & 1.30 & 0.66 & 107 & E-U-A \\ 
SDSSJ1041$+$0112 & 0274-51913-575 & 0.1006 & 0.2172 & 200 $\pm$    7 & 16.08 & 0.09 & 2.50 & 0.85 & 14 & E-S-A \\ 
\hline \hline
\end{tabular}
\end{center}
\textsc{      Note.} --- Column 1 is the SDSS system name. Column 2 provides a unique SDSS spectrum identifier. Columns 3 and 4 are the redshifts of the foreground lens and the background source inferred from the SDSS spectrum. Column 5 is the stellar velocity dispersion reported by the SDSS reduction pipeline. Column 6 provides the apparent AB magnitude of the lens galaxy in the F814W-band inferred from the de Vaucouleurs model. Galactic dust extinction values based on \citet{Schlegel98} maps are given in Column 7, and should be subtracted from the observed magnitude to give the dust-corrected magnitude. Columns 8, 9, and 10 are the effective radius (in the intermediate axis convention), minor-to-major axis ratio, and major-axis position angle of the lens galaxy with respect to the north inferred from \textsl{HST} F814W-band imaging data, assuming a de Vaucouleurs model. Column 11 is the classification with codes denoting the foreground-lens morphology, the foreground-lens multiplicity, and the status of the system as a lens based on the available data. Morphology is coded by ``E'' for early-type (elliptical and S0) and ``L'' for late-type (Sa and later). Multiplicity is coded by ``S'' for single and ``M'' for multiple. Lens status is coded by ``A'' for systems with clear and convincing evidence of multiple imaging, ``M'' for systems with possible evidence of multiple imaging, and ``X'' for nonlenses. \\
\end{table*}
\addtocounter{table}{-1}
\begin{table*}[htbp]
\begin{center}
\caption{\textit{Continued}}
\begin{tabular}{c c c c c c c c c c c}
\hline \hline
Target & Plate-MJD-Fiber ID & $z_{L}$ & $z_{S}$ & $\sigma_{\rm SDSS}$ & $I_{\rm 814}$ & $\Delta I_{\rm 814}$ & $R_{\rm eff}$ & $q$ & P.A. & Classification \\
 & & & & (km\,s$^{-1}$) & (mag) & (mag) & (arcsec) & & (deg) & \\
(1) & (2) & (3) & (4) & (5) & (6) & (7) & (8) & (9) & (10) & (11) \\
\hline
SDSSJ1048$+$1313 & 1749-53357-165 & 0.1330 & 0.6679 & 195 $\pm$   10 & 16.62 & 0.07 & 1.90 & 0.62 & 52 & E-S-A \\ 
SDSSJ1051$+$4439 & 1434-53053-142 & 0.1634 & 0.5380 & 216 $\pm$   16 & 17.06 & 0.03 & 1.66 & 0.78 & 15 & E-S-A \\ 
SDSSJ1056$+$4141 & 1362-53050-078 & 0.1343 & 0.8318 & 157 $\pm$   10 & 16.95 & 0.02 & 1.81 & 0.87 & 28 & E-S-A \\ 
SDSSJ1101$+$1523 & 2487-53852-203 & 0.1780 & 0.5169 & 270 $\pm$   15 & 17.22 & 0.04 & 0.89 & 0.71 & 32 & E-S-A \\ 
SDSSJ1102$+$3655 & 2091-53447-141 & 0.0937 & 0.1857 & 271 $\pm$    9 & 14.79 & 0.04 & 4.70 & 0.64 & 167 & E-S-X \\ 
SDSSJ1103$+$3625 & 2091-53447-101 & 0.1567 & 0.2655 & 282 $\pm$   14 & 15.77 & 0.04 & 2.77 & 0.73 & 135 & E-S-X \\ 
SDSSJ1116$+$0729 & 1617-53112-393 & 0.1697 & 0.6860 & 190 $\pm$   11 & 16.87 & 0.07 & 2.44 & 0.81 & 65 & E-S-A \\ 
SDSSJ1119$+$1357 & 1753-53383-269 & 0.0678 & 0.3851 & 206 $\pm$    5 & 14.88 & 0.05 & 4.37 & 0.65 & 80 & E-S-X \\ 
SDSSJ1121$+$5822 & 0951-52398-147 & 0.1751 & 0.3273 & 203 $\pm$   12 & 17.02 & 0.03 & 1.34 & 0.94 & 161 & E-S-X \\ 
SDSSJ1122$+$4748 & 1441-53083-526 & 0.1092 & 0.3451 & 112 $\pm$   12 & 16.71 & 0.03 & 4.42 & 0.59 & 136 & L-S-X \\ 
SDSSJ1127$+$2312 & 2497-54154-046 & 0.1303 & 0.3610 & 230 $\pm$    9 & 15.91 & 0.03 & 2.69 & 0.89 & 112 & E-S-A \\ 
SDSSJ1129$+$1953 & 2502-54180-383 & 0.1323 & 0.6981 & 229 $\pm$   15 & 16.93 & 0.05 & 1.45 & 0.71 & 131 & E-S-C \\ 
SDSSJ1137$+$1818 & 2503-53856-565 & 0.1241 & 0.4627 & 222 $\pm$    8 & 16.14 & 0.05 & 1.79 & 0.89 & 105 & E-S-A \\ 
SDSSJ1138$+$1134 & 1608-53138-306 & 0.1821 & 0.4773 & 194 $\pm$   13 & 17.01 & 0.07 & 1.60 & 0.76 & 124 & E-S-B \\ 
SDSSJ1142$+$2509 & 2505-53856-570 & 0.1640 & 0.6595 & 159 $\pm$   10 & 17.11 & 0.04 & 1.51 & 0.90 & 58 & E-S-A \\ 
SDSSJ1144$+$0436 & 0839-52373-230 & 0.1036 & 0.2551 & 207 $\pm$   14 & 16.97 & 0.04 & 1.22 & 0.83 & 173 & E-S-A \\ 
SDSSJ1148$+$3103 & 1991-53446-288 & 0.1425 & 0.2870 & 239 $\pm$   10 & 16.30 & 0.05 & 1.83 & 0.60 & 24 & E-S-X \\ 
SDSSJ1150$+$2944 & 2224-53815-277 & 0.2354 & 0.5710 & 223 $\pm$   14 & 16.55 & 0.04 & 2.86 & 0.77 & 125 & E-S-X \\ 
SDSSJ1203$+$1558 & 1764-53467-408 & 0.2649 & 0.4206 & 247 $\pm$   24 & 17.10 & 0.06 & 2.10 & 0.67 & 147 & E-S-B \\ 
SDSSJ1213$+$2930 & 2228-53818-064 & 0.0906 & 0.5954 & 232 $\pm$    7 & 15.82 & 0.04 & 1.73 & 0.67 & 70 & E-S-A \\ 
SDSSJ1229$+$1615 & 2598-54232-126 & 0.1207 & 0.7586 & 183 $\pm$   11 & 16.58 & 0.05 & 1.68 & 0.74 & 59 & E-S-X \\ 
SDSSJ1230$+$6519 & 0600-52317-496 & 0.1274 & 0.2725 & 191 $\pm$    9 & 16.70 & 0.04 & 1.63 & 0.87 & 43 & E-S-C \\ 
SDSSJ1235$+$3919 & 1984-53433-095 & 0.0623 & 0.1917 & 166 $\pm$    6 & 14.86 & 0.03 & 4.24 & 0.68 & 149 & E-S-X \\ 
SDSSJ1238$+$6709 & 0494-51915-074 & 0.2312 & 0.4447 & 223 $\pm$   10 & 16.40 & 0.04 & 6.58 & 0.62 & 122 & E-M-X \\ 
SDSSJ1242$+$0248 & 0521-52326-587 & 0.2056 & 0.8171 & 233 $\pm$   12 & 17.09 & 0.06 & 1.30 & 0.80 & 54 & E-S-X \\ 
SDSSJ1248$+$4458 & 1373-53063-432 & 0.2628 & 0.6706 & 236 $\pm$   23 & 17.08 & 0.05 & 2.88 & 0.83 & 157 & E-S-X \\ 
SDSSJ1258$+$3703 & 2018-53800-254 & 0.0733 & 0.4370 & 196 $\pm$    9 & 16.81 & 0.03 & 0.90 & 0.71 & 141 & E-S-B \\ 
SDSSJ1258$-$0259 & 0338-51694-221 & 0.1111 & 0.5068 & 151 $\pm$    9 & 16.86 & 0.05 & 1.56 & 0.45 & 65 & L-S-X \\ 
SDSSJ1301$+$0834 & 1793-53883-124 & 0.0902 & 0.5331 & 178 $\pm$    8 & 16.16 & 0.05 & 1.25 & 0.55 & 160 & E-S-A \\ 
SDSSJ1306$+$5545 & 1319-52791-287 & 0.0650 & 0.4872 & 142 $\pm$    8 & 15.96 & 0.03 & 1.78 & 0.97 & 90 & E-S-C \\ 
SDSSJ1310$+$0220 & 0525-52295-440 & 0.0665 & 0.5526 & 0 $\pm$   98 & 16.13 & 0.07 & 10.00 & 0.90 & 80 & E-S-C \\ 
SDSSJ1328$+$0156 & 0527-52342-181 & 0.1168 & 0.5068 & 154 $\pm$    8 & 16.30 & 0.05 & 1.89 & 0.72 & 86 & L-S-X \\ 
SDSSJ1330$+$1750 & 2641-54230-253 & 0.2074 & 0.3717 & 250 $\pm$   12 & 16.20 & 0.04 & 2.85 & 0.74 & 176 & E-S-A \\ 
SDSSJ1356$+$0601 & 1805-53875-017 & 0.1256 & 1.0882 & 0 $\pm$    0 & 16.59 & 0.05 & 3.26 & 0.90 & 28 & E-S-X \\ 
SDSSJ1400$+$2933 & 2122-54178-223 & 0.3407 & 0.8087 & 193 $\pm$   22 & 17.46 & 0.04 & 10.00 & 0.66 & 177 & E-S-X \\ 
SDSSJ1403$+$3309 & 2121-54180-444 & 0.0625 & 0.7720 & 190 $\pm$    6 & 15.56 & 0.03 & 2.00 & 0.81 & 51 & E-S-A \\ 
SDSSJ1405$+$1353 & 1704-53178-474 & 0.1331 & 0.2828 & 193 $\pm$   11 & 17.28 & 0.04 & 1.06 & 0.67 & 21 & E-S-X \\ 
SDSSJ1406$+$2528 & 2124-53770-362 & 0.1193 & 0.7285 & 406 $\pm$   17 & 16.96 & 0.04 & 1.47 & 1.00 & 149 & E-S-X \\ 
SDSSJ1422$+$0203 & 0534-51997-481 & 0.1104 & 0.5176 & 172 $\pm$    9 & 16.39 & 0.07 & 2.05 & 0.72 & 175 & E-S-C \\ 
SDSSJ1424$+$0109 & 0305-51613-510 & 0.3042 & 0.9287 & 327 $\pm$   27 & 16.56 & 0.06 & 5.19 & 0.75 & 47 & E-S-C \\ 
SDSSJ1425$+$0951 & 1707-53885-023 & 0.1583 & 0.4554 & 211 $\pm$   11 & 16.88 & 0.05 & 1.14 & 0.74 & 72 & E-S-C \\ 
SDSSJ1430$+$6104 & 0607-52368-404 & 0.1688 & 0.6537 & 180 $\pm$   15 & 16.72 & 0.02 & 2.24 & 0.79 & 160 & E-S-A \\ 
SDSSJ1433$+$2835 & 2134-53876-575 & 0.0912 & 0.4115 & 230 $\pm$    6 & 15.17 & 0.03 & 3.23 & 0.95 & 104 & E-S-A \\ 
SDSSJ1446$+$4943 & 1047-52733-508 & 0.1731 & 0.3414 & 214 $\pm$   12 & 16.98 & 0.05 & 1.64 & 0.91 & 174 & E-S-C \\ 
SDSSJ1451$+$2951 & 2141-53764-597 & 0.1249 & 0.2687 & 245 $\pm$    8 & 15.83 & 0.03 & 2.53 & 0.74 & 169 & E-S-X \\ 
SDSSJ1514$+$3003 & 1845-54144-573 & 0.0923 & 0.6977 & 189 $\pm$    7 & 15.80 & 0.05 & 2.43 & 0.82 & 70 & E-S-X \\ 
SDSSJ1514$+$3259 & 1386-53116-225 & 0.1124 & 0.7154 & 203 $\pm$    9 & 16.72 & 0.03 & 1.55 & 0.62 & 25 & E-S-X \\ 
SDSSJ1531$+$0652 & 1820-54208-391 & 0.2085 & 0.2959 & 265 $\pm$   15 & 16.40 & 0.08 & 4.19 & 0.83 & 147 & E-U-C \\ 
SDSSJ1541$+$3642 & 1416-52875-381 & 0.1406 & 0.7389 & 194 $\pm$   11 & 16.57 & 0.04 & 1.55 & 0.94 & 142 & E-S-A \\ 
SDSSJ1542$+$3214 & 1581-53149-173 & 0.0924 & 0.3510 & 174 $\pm$   10 & 16.02 & 0.06 & 3.22 & 0.91 & 63 & E-S-X \\ 
SDSSJ1543$+$2202 & 2166-54232-606 & 0.2681 & 0.3966 & 285 $\pm$   16 & 16.90 & 0.11 & 2.32 & 0.80 & 11 & E-S-A \\ 
SDSSJ1550$+$2020 & 2168-53886-595 & 0.1351 & 0.3501 & 243 $\pm$    9 & 16.29 & 0.10 & 1.68 & 0.68 & 133 & E-S-A \\ 
SDSSJ1553$+$3004 & 1579-53473-235 & 0.1604 & 0.5663 & 194 $\pm$   15 & 17.05 & 0.06 & 2.15 & 0.92 & 78 & E-S-A \\ 
SDSSJ1607$+$1545 & 2197-53555-065 & 0.1422 & 0.4105 & 167 $\pm$   14 & 16.96 & 0.08 & 2.09 & 0.97 & 71 & E-S-X \\ 
SDSSJ1607$+$2147 & 2205-53793-414 & 0.2089 & 0.4865 & 197 $\pm$   16 & 17.14 & 0.16 & 2.63 & 0.90 & 45 & E-S-A \\ 
SDSSJ1609$+$1805 & 2200-53875-568 & 0.1497 & 0.5222 & 225 $\pm$   10 & 16.38 & 0.09 & 2.18 & 0.78 & 74 & E-S-X \\ 
SDSSJ1610$+$4648 & 0813-52354-071 & 0.0462 & 0.3028 & 48 $\pm$   28 & 17.03 & 0.02 & 10.00 & 0.83 & 48 & U-S-X \\ 
SDSSJ1621$+$0552 & 1731-53884-010 & 0.1538 & 0.4203 & 193 $\pm$   21 & 17.14 & 0.12 & 1.29 & 0.85 & 110 & E-U-C \\ 
SDSSJ1629$+$1331 & 2204-53877-356 & 0.1223 & 1.2196 & 176 $\pm$    9 & 16.84 & 0.09 & 1.39 & 0.72 & 40 & E-S-X \\ 
SDSSJ1633$+$1441 & 2204-53877-379 & 0.1281 & 0.5804 & 231 $\pm$    9 & 16.04 & 0.11 & 2.39 & 0.83 & 113 & E-S-A \\ 
SDSSJ2134$+$1043 & 0731-52460-165 & 0.2290 & 0.3963 & 240 $\pm$   14 & 16.33 & 0.12 & 3.43 & 0.89 & 144 & E-S-X \\ 
SDSSJ2157$+$0004 & 0372-52173-437 & 0.1444 & 0.3414 & 176 $\pm$   14 & 16.87 & 0.11 & 1.86 & 0.67 & 164 & E-S-X \\ 
SDSSJ2211$-$0843 & 0718-52206-091 & 0.0684 & 0.7277 & 139 $\pm$    6 & 16.12 & 0.10 & 2.16 & 0.79 & 62 & E-S-C \\ 
SDSSJ2309$-$0039 & 0381-51811-163 & 0.2905 & 1.0048 & 184 $\pm$   13 & 17.29 & 0.07 & 2.08 & 0.96 & 107 & E-S-A \\ 
SDSSJ2324$+$0105 & 0680-52200-564 & 0.1899 & 0.2775 & 245 $\pm$   15 & 17.19 & 0.08 & 1.10 & 0.53 & 54 & E-S-A \\ 
SDSSJ2356$+$1427 & 0749-52226-067 & 0.1446 & 0.2673 & 204 $\pm$   14 & 16.32 & 0.08 & 2.61 & 0.65 & 96 & E-S-X \\ 
\hline \hline
\end{tabular}
\end{center}
\end{table*}

\begin{table*}[htbp]
\begin{center}
\caption{\label{tb:tb2} Strong-lens model parameters of the 40 S4TM grade-A lenses.}
\begin{tabular}{c c c c c c c c c c}
\hline \hline
Target & $b_{\rm SIE}$ & $q$ & P.A. & $N_{\rm source}$ & $\mu$ & $\log_{10} (M_{\rm Ein}/M_{\odot})$ & $\log_{10} (M_*^{\rm Chab}/M_{\odot})$ & $f_{\rm dm}$ & $\chi^2$/dof \\
 & (arcsec) & & (deg) & & & & & & \\
(1) & (2) & (3) & (4) & (5) & (6) & (7) & (8) & (9) & (10)\\
\hline
SDSSJ0143$-$1006 & 1.23 & 0.64 & 75 & 1 & 3 & 11.26 & 11.53 & 0.49 & 30569./24175 \\ 
SDSSJ0159$-$0006 & 0.92 & 0.75 & 114 & 1 & 6 & 10.89 & 11.03 & 0.56 & 18137./24453 \\ 
SDSSJ0324$+$0045 & 0.55 & 0.82 & 20 & 1 & 14 & 10.79 & 11.31 & 0.02 & 23713./13727 \\ 
SDSSJ0324$-$0110 & 0.63 & 0.47 & 83 & 1 & 4 & 11.36 & 11.71 & 0.52 & 14108./13293 \\ 
SDSSJ0753$+$3416 & 1.23 & 0.87 & 141 & 4 & 24 & 11.05 & 11.23 & 0.42 & 37313./13799 \\ 
SDSSJ0754$+$1927 & 1.04 & 0.73 & 26 & 1 & 6 & 10.99 & 11.13 & 0.33 & 22166./19148 \\ 
SDSSJ0757$+$1956 & 1.62 & 0.85 & 133 & 2 & 9 & 11.24 & 11.34 & 0.61 & 28086./24187 \\ 
SDSSJ0826$+$5630 & 1.01 & 0.96 & 82 & 1 & 105 & 10.85 & 11.38 & 0.09 & 21812./12732 \\ 
SDSSJ0847$+$2348 & 0.96 & 0.94 & 70 & 2 & 17 & 10.97 & 11.19 & 0.44 & 24039./18714 \\ 
SDSSJ0851$+$0505 & 0.91 & 0.87 & 53 & 3 & 6 & 10.79 & 11.05 & 0.23 & 17546./13802 \\ 
SDSSJ0920$+$3028 & 0.70 & 0.88 & 86 & 1 & 8 & 11.34 & 12.08 & 0.39 & 10811./9356 \\ 
SDSSJ0955$+$3014 & 0.54 & 0.82 & 161 & 1 & 7 & 11.08 & 11.77 & 0.38 & 10066./9743 \\ 
SDSSJ0956$+$5539 & 1.17 & 0.96 & 88 & 1 & 19 & 11.19 & 11.46 & 0.32 & 17705./13764 \\ 
SDSSJ1010$+$3124 & 1.14 & 0.65 & 78 & 1 & 4 & 11.21 & 11.68 & 0.45 & 16668./18966 \\ 
SDSSJ1031$+$3026 & 0.88 & 0.70 & 9 & 3 & 5 & 10.88 & 11.22 & -0.16 & 19210./13772 \\ 
SDSSJ1040$+$3626 & 0.59 & 0.88 & 95 & 2 & 3 & 10.54 & 10.99 & 0.33 & 18880./13512 \\ 
SDSSJ1041$+$0112 & 0.60 & 0.87 & 52 & 2 & 5 & 10.50 & 11.07 & 0.39 & 14837./13968 \\ 
SDSSJ1048$+$1313 & 1.18 & 0.64 & 49 & 3 & 4 & 11.03 & 11.22 & 0.52 & 12426./14109 \\ 
SDSSJ1051$+$4439 & 0.99 & 0.76 & 21 & 1 & 3 & 11.02 & 11.16 & 0.42 & 20182./18441 \\ 
SDSSJ1056$+$4141 & 0.72 & 0.79 & 55 & 1 & 10 & 10.59 & 11.12 & 0.35 & 16193./13774 \\ 
SDSSJ1101$+$1523 & 1.18 & 0.81 & 20 & 1 & 5 & 11.23 & 11.23 & 0.25 & 15033./13542 \\ 
SDSSJ1116$+$0729 & 0.82 & 0.85 & 144 & 1 & 4 & 10.83 & 11.29 & 0.36 & 16934./12512 \\ 
SDSSJ1127$+$2312 & 1.25 & 0.90 & 111 & 1 & 8 & 11.18 & 11.44 & 0.50 & 20505./18858 \\ 
SDSSJ1137$+$1818 & 1.29 & 0.89 & 114 & 1 & 10 & 11.12 & 11.31 & 0.40 & 15057./13832 \\ 
SDSSJ1142$+$2509 & 0.79 & 0.80 & 0 & 1 & 18 & 10.80 & 11.21 & 0.28 & 16638./13878 \\ 
SDSSJ1144$+$0436 & 0.76 & 0.79 & 119 & 1 & 5 & 10.68 & 10.74 & 0.48 & 18128./13840 \\ 
SDSSJ1213$+$2930 & 1.35 & 0.75 & 72 & 1 & 21 & 10.98 & 11.09 & 0.34 & 19766./13880 \\ 
SDSSJ1301$+$0834 & 1.00 & 0.78 & 157 & 2 & 9 & 10.72 & 10.92 & 0.05 & 11690./13727 \\ 
SDSSJ1330$+$1750 & 1.01 & 0.78 & 14 & 1 & 4 & 11.32 & 11.74 & 0.37 & 20408./24398 \\ 
SDSSJ1403$+$3309 & 1.02 & 0.85 & 54 & 1 & 9 & 10.55 & 10.78 & 0.28 & 6631./14091 \\ 
SDSSJ1430$+$6104 & 1.00 & 0.75 & 161 & 2 & 11 & 11.01 & 11.32 & 0.35 & 12764./13463 \\ 
SDSSJ1433$+$2835 & 1.53 & 0.91 & 120 & 1 & 10 & 11.12 & 11.45 & 0.55 & 14345./24778 \\ 
SDSSJ1541$+$3642 & 1.17 & 0.91 & 74 & 1 & 16 & 11.04 & 11.25 & 0.29 & 19801./18549 \\ 
SDSSJ1543$+$2202 & 0.78 & 0.72 & 12 & 1 & 3 & 11.32 & 11.74 & 0.45 & 19060./13243 \\ 
SDSSJ1550$+$2020 & 1.01 & 0.71 & 146 & 2 & 2 & 11.02 & 11.30 & 0.26 & 22011./24139 \\ 
SDSSJ1553$+$3004 & 0.84 & 0.83 & 59 & 1 & 5 & 10.86 & 11.26 & 0.53 & 15143./13733 \\ 
SDSSJ1607$+$2147 & 0.57 & 0.57 & 169 & 1 & 2 & 10.71 & 11.55 & 0.50 & 15809./13643 \\ 
SDSSJ1633$+$1441 & 1.39 & 0.93 & 115 & 2 & 26 & 11.17 & 11.39 & 0.47 & 6765./13026 \\ 
SDSSJ2309$-$0039 & 1.14 & 0.89 & 41 & 1 & 4 & 11.35 & 11.68 & 0.27 & 28622./17981 \\ 
SDSSJ2324$+$0105 & 0.59 & 0.98 & 113 & 1 & 8 & 10.97 & 11.32 & 0.35 & 16402./9725 \\ 
\hline \hline
\end{tabular}
\end{center}
\textsc{      Note.} --- Column 1 is the SDSS system name. Columns 2--4 are the Einstein radius, minor-to-major axis ratio, and major-axis position angle of the SIE component with respect to the north. Column 5 indicates the number of S\'{e}rsic components used. Column 6 is the average magnification. Column 7 is the total projected mass within the Einstein radius from the best-fit lens model. Column 8 is the estimated stellar mass assuming a Chabrier IMF from \citet{Shu15}. Column 9 is the inferred dark-matter fraction within half of the half-light radius. Column 10 provides the $\chi^2$ value and the dof. \\
\end{table*}

\begin{figure*}[htbp]
\centerline{\plotone{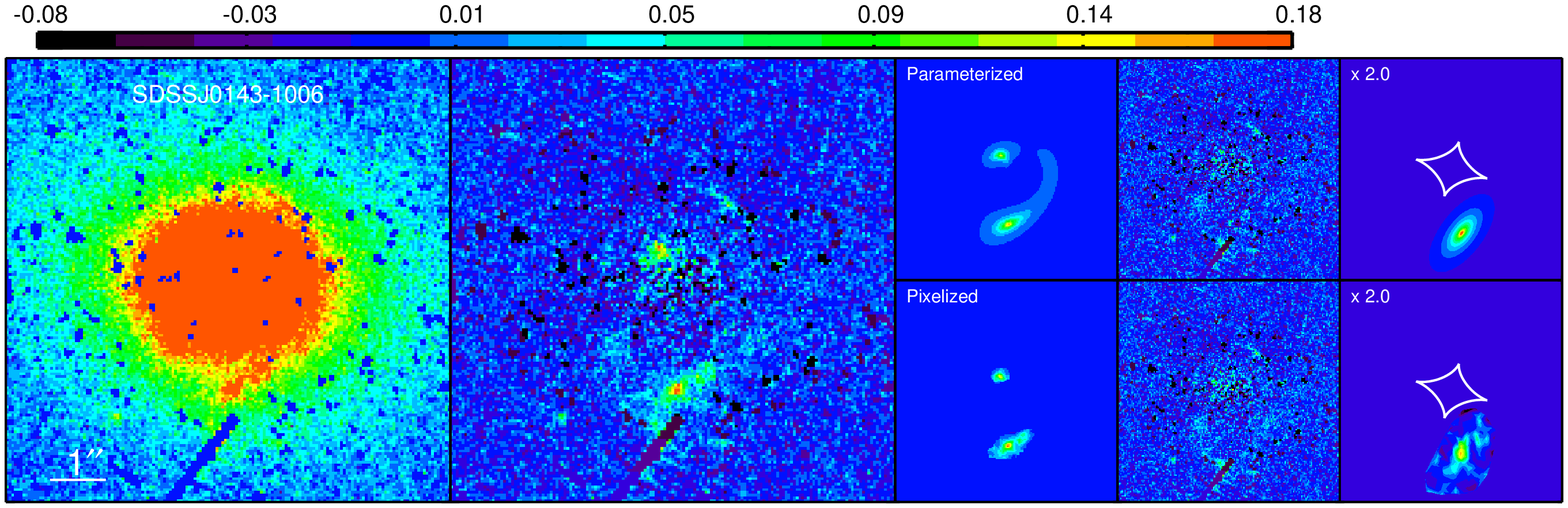}}
\centerline{\plotone{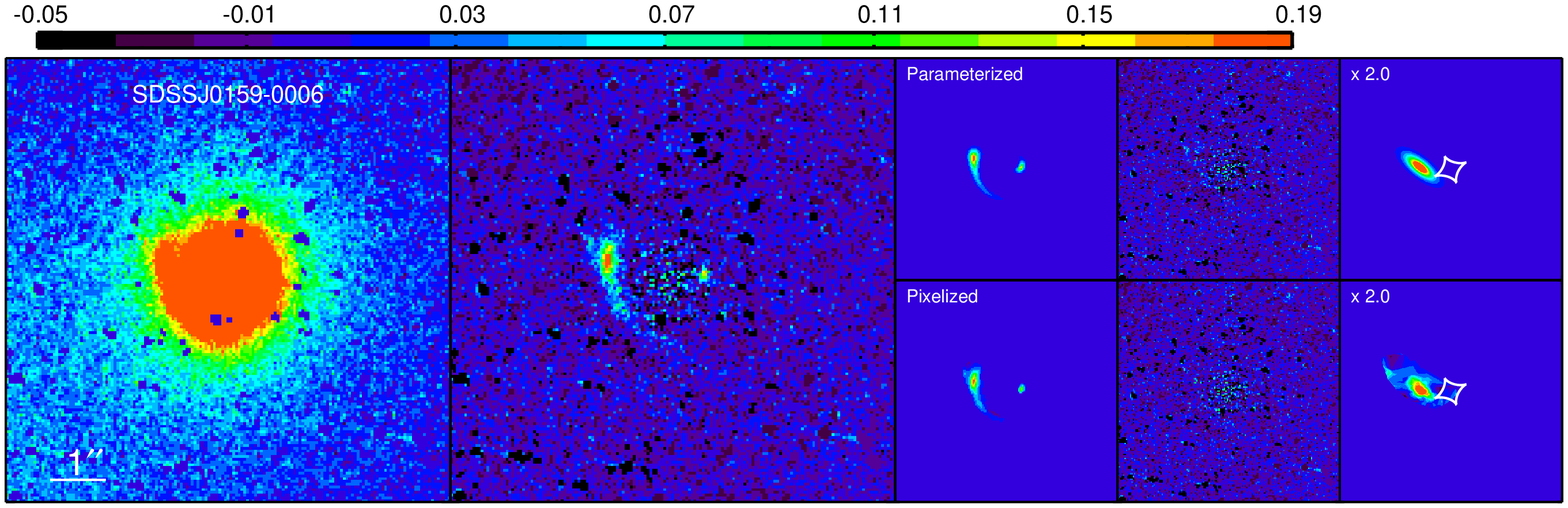}}
\centerline{\plotone{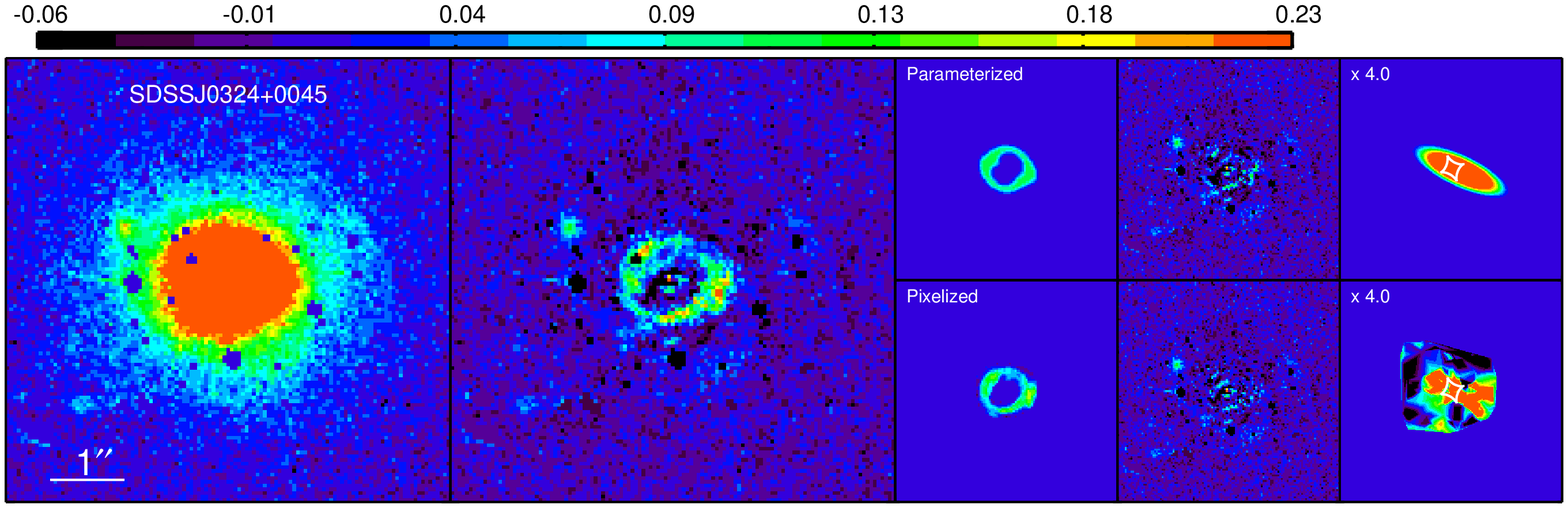}}
\centerline{\plotone{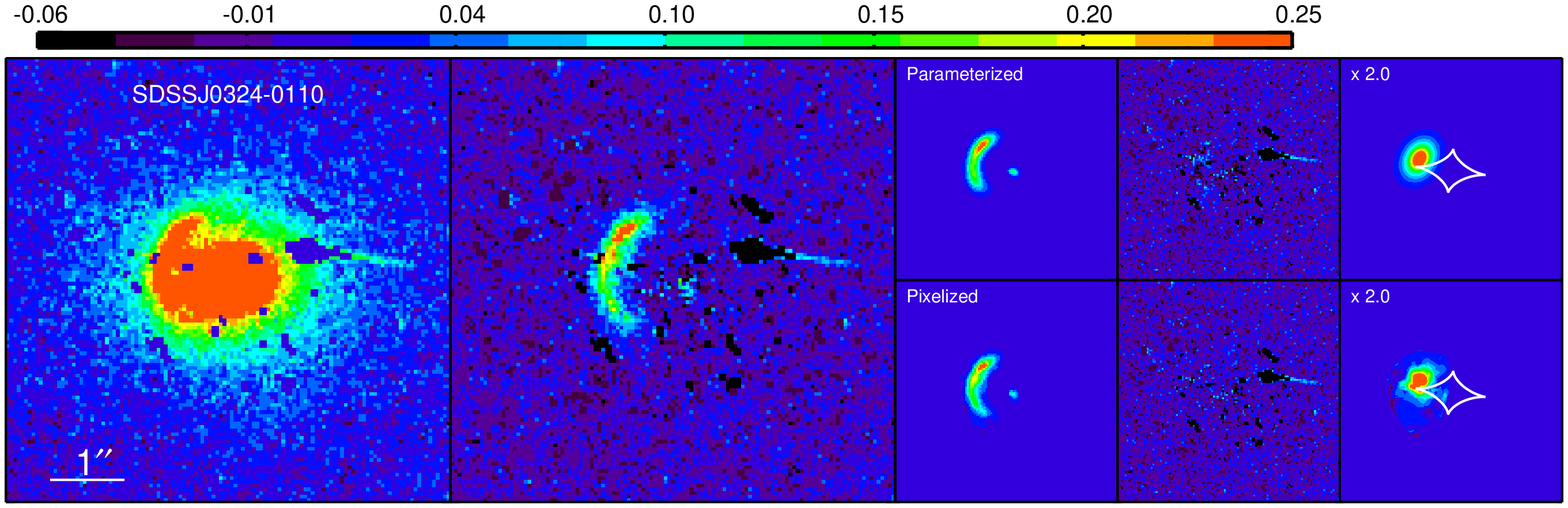}}
\caption{\label{fig:models_1} SIE lens models for the 40 S4TM grade-A lenses. The observational data, B-spline-subtracted image, predicted lensed image, final residual, and the background source model are shown from left to right, respectively. Images are orientated such that north is up and East is to the left. For each system, the results of the two source models are split into two rows with the parameterized source model on the top and the pixelized source model on the bottom. The white lines in the last panels are the caustics of the lens model. The source plane panels are magnified by factors from 2 to 32 relative to the image plane panel as indicated in each panel. The color bars indicate the intensity levels in units of electrons per second per pixel$^2$. [\textit{The remaining 36 figures are available in the online journal.}]}
\end{figure*}

Table~\ref{tb:tb2} lists the best-fit parameters for the 40 S4TM grade-A lenses including 
the Einstein radius $b_{\rm SIE}$, the minor-to-major axis ratio $q$, and the major-axis 
position angle P.A. of the SIE model, the number of S\'{e}rsic components $N_{\rm source}$, 
$\chi^2$ value, the degree of freedom (dof), and the average magnification $\mu$ defined 
as the ratio of the total flux mapped onto the image plane to the total flux in the 
source plane. From the best-fit lens models for the 40 grade-A lenses shown in 
Figure~\ref{fig:models_1}, it can be seen that the simple SIE model provides satisfactory
fits to the observational data. The background sources are typically resolved into 1--4 
clumps with a typical average magnification of 7. 
The average and median values of the reduced $\chi^2$ are 1.18 and 1.16, respectively. 
This again confirms that external shear is negligible for these lens systems. 
Benefited from strong lensing, we can infer the total projected mass within the 
Einstein radius of each lens galaxy, $M_{\rm Ein}$, as 
\begin{equation}
M_{\rm Ein} = \pi (b_{\rm SIE} d_L)^2 \, \Sigma_{\rm crit}. 
\end{equation}
\citet{Shu15} derived the stellar masses $M_*^{\rm Chab}$ of all the S4TM lens galaxies 
based on their \textsl{HST} F814W-band photometric data and a simple stellar population 
synthesis model assuming a Chabrier initial mass function \citep[IMF;][]{Chabrier03}, and 
further calculated the projected dark-matter fraction within one half of the 
half-light radius $f_{\rm dm}$. These values are also reported in Table~\ref{tb:tb2}. 
As shown in \citet{Shu15}, a strong trend of increasing dark-matter fraction at higher 
galaxy mass is detected. 

\section{Discussion}
\label{sect:discussion}

\begin{figure*}
\centering
\plotone{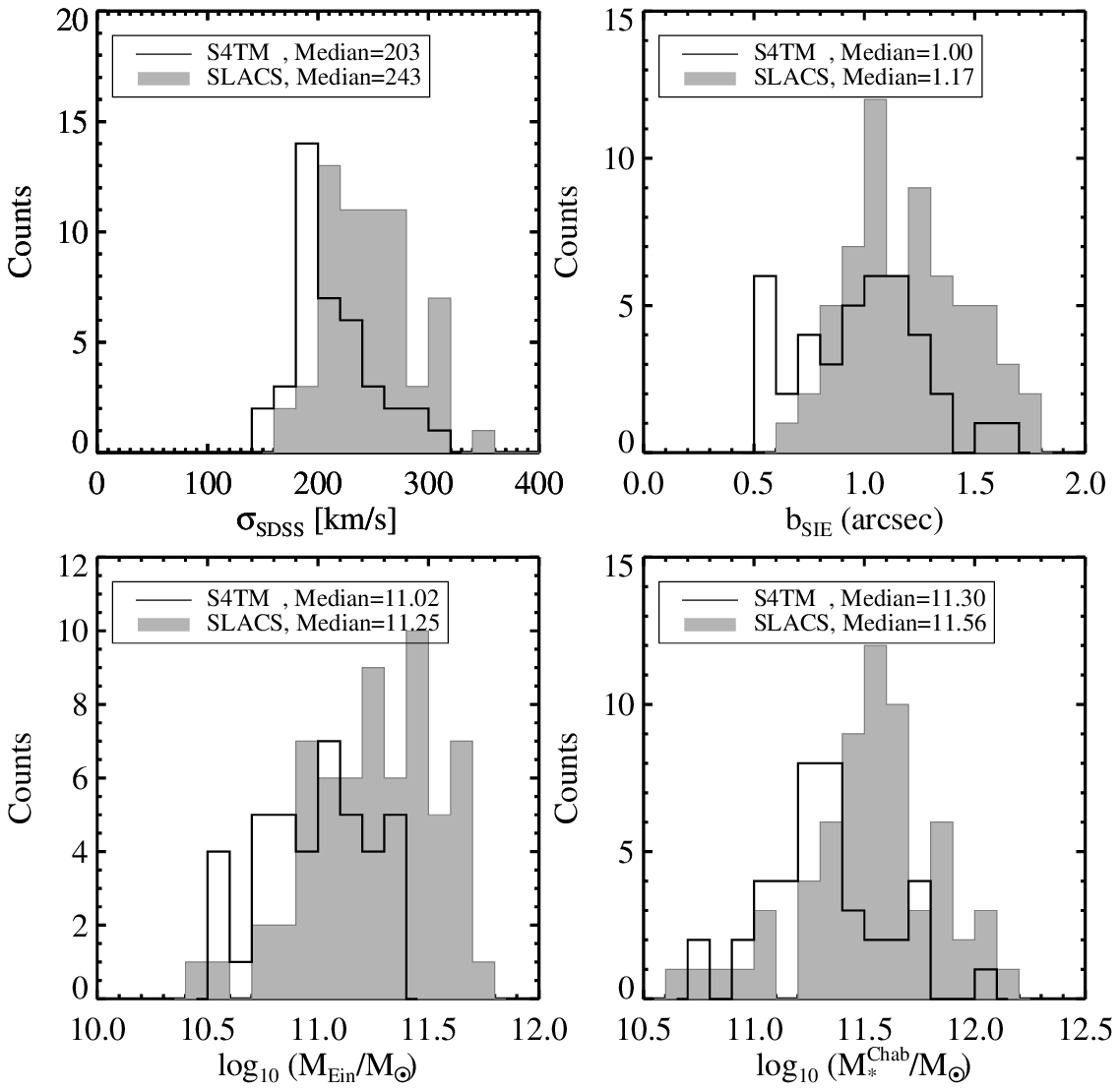}
\caption{\label{fig:histograms}
Distributions of the stellar velocity dispersion, Einstein radius, 
total enclosed mass within the Einstein radius, and stellar mass for the S4TM 
(solid histograms) and the SLACS (dashed histograms) lens samples. The stellar mass is 
derived from \textsl{HST} F814W-band photometry assuming a Chabrier IMF as explained 
in \citet{Shu15}.}
\end{figure*}

The S4TM survey is optimized to select strong-lens systems with relatively lower-mass lens 
galaxies as a complementary sample to the SLACS sample. 
In Figure~\ref{fig:histograms}, we compare the stellar velocity dispersion, Einstein radius, 
total projected mass within the Einstein radius, and stellar mass 
between the two lens samples. The SLACS sample refers to the 63 grade-A lenses 
with lens models in \citet{SLACSV}. 

The distribution of the stellar velocity dispersions 
$\sigma_{\rm SDSS}$ of the S4TM sample has a strong peak at about 200 km\,s$^{-1}$ and 
declines rapidly on both ends. On the other hand, the distribution of $\sigma_{\rm SDSS}$ 
of the SLACS sample is almost flat from 200 to 320 km\,s$^{-1}$. 
As a comparison, the median $\sigma_{\rm SDSS}$ of the S4TM sample is 203 km\,s$^{-1}$ while 
it is 243 km\,s$^{-1}$ for the SLACS sample. 
The median Einstein radius of the S4TM sample is 1\farcs00, 
almost $15\%$ smaller than that of the SLACS sample (1\farcs17). 
The distributions of the Einstein radii for the two samples further suggest that 
the S4TM sample is more abundant in systems with Einstein radii smaller than 0\farcs8 
(12/40 versus 3/63) and lack systems with Einstein radii larger than 1\farcs2 
(8/40 versus 30/63). 
We basically expect the distributions of $M_{\rm Ein}$ to be similar to those of $b_{\rm SIE}$ 
because $\Sigma_{\rm crit}$, which is determined by the lens and source redshifts, 
distributes roughly the same for the two samples. The histogram in Figure~\ref{fig:histograms} 
confirms this. The total projected mass within the Einstein radius $M_{\rm Ein}$ of the S4TM 
sample ranges from $3 \times 10^{10}$ to $2 \times 10^{11} M_{\odot}$. 
And the median $\log_{10} (M_{\rm Ein}/M_{\odot})$ of the S4TM sample is 11.02, 
0.23 dex (almost a factor of 2) smaller than that of the SLACS lens galaxies. 
The stellar mass for these lens galaxies ranges from $3\times10^{10}$ to 
$1\times10^{12} M_{\odot}$. 
The S4TM grade-A lens galaxies are again less massive in stellar mass than 
SLACS grade-A lens galaxies. The difference in the median values is 0.26 dex. 
We also look at the ratio of the Einstein radius to the half-light radius for these two lens 
samples. The distributions appear almost the same and peak around 0.5. The median 
$b_{\rm SIE}/r_{\rm half}$ ratio of the S4TM sample is slightly larger than that of the 
SLACS sample (0.54 versus 0.48). 

The comparatively less massive S4TM lens sample serves as a complementary addition to 
the current galaxy-scale strong-lens samples, which are usually biased toward 
more massive lens galaxies \citep[e.g.,][]{SLACSX, Faure11, Brownstein12, Sonnenfeld13}. 
It extends the lens-galaxy mass coverage to the lower-mass end, and can allow a more 
thorough investigation of the mass structure and scaling relations of ETGs 
when combined with other strong-lens samples, especially the SLACS sample 
which is selected from the same parent sample with the same selection technique. 
For instance, we studied the mass structure of ETGs by combining S4TM and SLACS 
grade-A and grade-C lenses. Previous studies with only high-mass coverages showed that 
the total mass--density distribution of ETGs in strong-lens systems can be well approximated 
by an isothermal profile with little correlation with galaxy mass 
\citep[e.g.,][]{SLACSIII, SLACSVII, Koopmans09, Barnabe11, Ruff11}. 
However, by including the relatively lower-mass S4TM grade-A lenses and also grade-C lenses, 
we found the total mass--density profile of ETGs varies systematically with galaxy mass 
with a 6$\sigma$ significance \citep{Shu15}. 

Although the S4TM sample does not reach as 
low as the lower characteristic mass scale of $3 \times10^{10} M_{\odot}$ in stellar mass, 
the broader mass coverage can still enable us to directly test, with the aid of 
strong lensing, for a transition in structural and dark-matter content trends at  
intermediate galaxy mass as noticed in previous studies 
\citep[e.g.,][]{Tremblay96, Graham03a, vanderWel09, Bernardi11a, Bernardi11b, 
Cappellari13a, Cappellari13}. Furthermore, the S4TM strong-lens sample can be a 
useful resource for testing general relativity (GR) by comparing dynamical mass and 
lensing mass \citep[e.g.,][]{Bolton06c, Jain08, Schwab10, Cao17}. In particular, 
the extended mass coverage of the S4TM sample will provide extra constraints 
on GR by revealing the environmental dependence of dark-matter halo properties as 
demonstrated by the numerical simulations \citep[e.g.,][]{Zhao11, Winther12, He14}. 
Lastly, we note that by further going to candidates with lower predicted lensing 
cross sections, we might be able to obtain a sample of strong-lens systems with 
even lower lens masses. 

\section{Summary}
\label{sect:summary}

In this paper, we presented a catalog of 40 new galaxy-scale strong lenses confirmed by 
\textsl{HST} F814W-band imaging observations of 118 candidates in the S4TM survey, 
an extension of the SLACS survey toward lower lens-galaxy mass. 
The \textsl{HST} observational data are well explained by an elliptical B-spline model 
for the lens-light distribution, an SIE profile for the lens-mass distribution, and 
multiple S\'{e}rsic components for the source-light distribution. Our main findings are as follows. 
\begin{enumerate}
\item The lens galaxies are ETGs at redshifts of $\sim 0.17$, and background sources are 
star-forming galaxies located at redshifts of $\sim 0.61$ with strong nebular emission lines 
(Balmer series, [O\textsc{ii}]~3727, or [O\textsc{iii}]~5007).
\item The Einstein radius distribution of the S4TM lenses ranges from 0\farcs54 to 1\farcs62 
with a median value of 1\farcs00. 
The fraction of systems with small Einstein radii ($<$ 0\farcs80) in the S4TM sample is 
a factor of 5 larger than that in the SLACS sample.
\item On average, the S4TM lenses are indeed less massive than those of the SLACS lenses. 
Based on our best-fit lens models, the total projected mass within the Einstein radius of 
the S4TM sample ranges from $3 \times 10^{10}$ to $2 \times 10^{11} M_{\odot}$ with 
a median mass of $1 \times 10^{11} M_{\odot}$, which is smaller by almost a factor of 2 
when compared to the SLACS sample. The SPS-derived stellar mass based on 
\textsl{HST} photometry also suggests that S4TM lenses are generally less massive than 
SLACS lenses by almost a factor of 2.
\item The extended mass coverage toward the low-mass end provided by the S4TM sample makes 
it a complementary addition to the current galaxy-scale strong-lens samples, and will also 
extend our understanding of ETGs. \citet{Shu15}, by including the relatively less massive 
S4TM grade-A lenses and grade-C lenses, detected a strong correlation between ETG mass and 
its total mass--density profile, which was not noticed in previous studies using only massive 
ETGs \citep[e.g.,][]{SLACSVII, Koopmans09, Barnabe11, Ruff11}. 
In addition, it enables us to probe intermediate-mass ETGs where transitions in scaling 
relations, kinematic properties, mass structure, and dark-matter content trends are detected 
\citep[e.g.,][]{Tremblay96, Graham03a, Kauffmann03, Graham08, Hyde09, Skelton09, Tortora09, 
vanderWel09, Bernardi11a, Bernardi11b, Cappellari13a, Cappellari13, Montero16}. 
\end{enumerate}

\acknowledgments

We thank the anonymous referee for helpful comments. 
Y.S. has been partially supported by the 973 program (No. 2015CB857003) and 
the National Natural Science Foundation of China (NSFC) 
under grant numbers 11603032 and 11333008. 
L.V.E.K is supported through an NWO-VICI grant (project number 639.043.308). 
T.T. acknowledges support from the Packard Foundation through a Packard Research Fellowship. 
A.M.D. acknowledges support from the Funda\c{c}\~ao de Amparo \`a Pesquisa do 
Estado de S\~ao Paulo (FAPESP), through the grant 2016/23567--4. 
R.G. acknowledges support for the Centre National des Etudes Spatiales. The work of L.A.M. was carried out at Jet Propulsion Laboratory, California Institute of Technology, under a contract with NASA. 

Support for Program \#12210 was provided by NASA through a grant from the Space Telescope Science Institute, which is operated by the Association of Universities for Research in Astronomy, Inc., under NASA contract NAS 5-26555.


\end{document}